\documentclass[preprint,secnumarabic,amssymb, aps, prl]{revtex4-1}
\usepackage{amssymb, latexsym}
\usepackage{amsmath} 
\usepackage{amsthm}
\usepackage{bm}
\usepackage{booktabs}
\usepackage[mathscr]{eucal}
\usepackage{enumerate}
\usepackage{url}
\usepackage{graphicx}
\usepackage{booktabs}
\usepackage{lineno}
 \usepackage{color}
\begin{document} 

\title{Enhanced sensing and conversion of ultrasonic Rayleigh waves by elastic metasurfaces}

\author{Andrea Colombi}%
\email[Corresponding author ]{e-mail: andree.colombi@gmail.com}
\author{Richard V. Craster}
\affiliation{Dept. of Mathematics, Imperial College London, South Kensington Campus, London}
\author{Victoria Ageeva}
\author{Richard J. Smith}
\author{Rikesh Patel}
\author{Matt Clark}
\affiliation{Optics and Photonics, University of Nottingham, U.K.}
\author{Daniel Colquitt}
\affiliation{Dept. of Mathematical Sciences, University of Liverpool, U.K.}
\author{Adam Clare}
\affiliation{Dept. of Mechanical, Materials and Manufacturing Engineering, University of Nottingham, U.K.}
\author{Sebastien Guenneau}
\affiliation{Institut Fresnel-CNRS (UMR 7249), Aix-Marseille Universit\'e, 13397 Marseille cedex 20, France}
\author{Philippe Roux}%
\affiliation{ISTerre, CNRS, Univ. Grenoble Alpes, France, BP 53 38041 Grenoble CEDEX 9}

\date{\today}

%%%%%%%%%%%%%%%%% END OF PREAMBLE %%%%%%%%%%%%%%%%

\begin{abstract}
Recent years have heralded the introduction of metasurfaces that advantageously combine the vision of sub-wavelength wave manipulation, with the design, fabrication and size advantages associated with surface excitation. 
An important topic within metasurfaces is the tailored rainbow trapping and selective spatial frequency separation of electromagnetic and acoustic waves using graded metasurfaces. This frequency dependent trapping and spatial frequency segregation has implications for energy concentrators and associated energy harvesting, sensing and wave filtering techniques. Different demonstrations of acoustic and electromagnetic rainbow devices have been performed, however not for deep elastic substrates that support both shear and compressional waves, together with surface Rayleigh waves; these allow not only for Rayleigh wave rainbow effects to exist but also for mode conversion from surface into bulk waves. We design, and test experimentally, a device creating not only elastic Rayleigh wave rainbow trapping, but also selective mode conversion of surface Rayleigh waves to shear waves. 

%The graded metasurfaces, the metawedge we design, are not limited to guided ultrasonic waves and are a general phenomenon in elastic waves that can be translated across scales. 

\end{abstract}

\maketitle
\noindent
Elastic rainbow phenomena, capable of spatially selecting surface wave packets, based upon frequency, have recently been demonstrated numerically in connection to seismic metamaterials \cite{colombi16a}. In contrast to acoustic \cite{acou_rain,romero13a} and electromagnetic \cite{hess2007} situations, where the main advantage of the rainbow effect is the spatial segregation of waves accompanied by the local enhancement of their amplitude, the elastic rainbow for elastic surface waves has an additional remarkable property which is that of broadband mode conversion from Rayleigh surface waves into bulk shear waves. 

Rainbow devices combine the remarkable properties and capabilities of locally resonant metamaterials, such as subwavelength bandgaps and trapping \cite{li,matthieu,andrea4,kaina}, with convenient graded or chirped structures. 
Graded designs take advantage of slow sound (or slow light), generated by spatial graduation, progressively decreasing the group velocity of waves they support. This creates media with spatially varying refraction index that find wide application in phononics (or photonics) because of their excellent abilities to manipulate and filter waves in compact devices \cite{yu14a,maigyte13a,kadic11a,christensen12a,lens_nature,andrea6}. Wave control with a graded design does not always rely on resonant phenomena; spatially varying index material for, say, gradient index (GRIN) lenses \cite{tyc2009} can alternatively, and more easily, be obtained through graded composite media \cite{yan2013,torrent_plate2,andrea_soil_lenses}. However the use of local resonators, as here, gives access to the very subwavelength scale and the power to have precise control over long waves.  

For instance, in rainbow trapping, the combined graded and resonant structure allows waves to be slowed down selectively upon their frequency, and eventually to be trapped in a subwavelength area, producing a strong local enhancement of their amplitude \cite{hess2007,acou_rain,romero13a,tian17a}. 
This enhancement is attracting most of the attention as it can positively impact technologies for energy harvesting, sensing and absorption of waves in electromagnetism \cite{jang11a,gan2009}, acoustics\cite{jimenez16a,sensing}, and, as discussed here, elastodynamics \cite{colombi16a}.

%would cut this... I would stick to elasticity/acoustics

%These ideas are closely related to using waveguides containing tapered metamaterial cores of negative refractive material. Combining ideas from metamaterials with explicitly subwavelength resonators, slow light, and tapering waveguides, a trapped rainbow of light was constructed\cite{hess2007} using a transparent dielectric medium containing tiny metallic subwavelength resonant inclusions of various shapes and arrangements. The physical description being that as light passes through this plasmonic wedge, the  oscillating electric currents generate electromagnetic field moments to trap light at a critical point at which the effective thickness of the plasmonic waveguide reduces to zero. This combined with the optical properties of surface plasmon polaritons (SPPs) \cite{pendry2004} has led to the development of surface dispersion engineering devices often based upon chirped gratings \cite{gan11a} that have application to ....
%more modern references

Despite the success of these ideas, there have been few applications of chirped and graded resonant structures to mechanical metamaterial devices. However, dramatic improvements in fabrication techniques (additive manufacturing, self-assembly and 3D printing), mean that the complexity of smoothly varying designs can now be easily realised. For instance, only recently this has been considered for GRIN lenses  \cite{tyc2009} for flexural waves using resonators on an elastic plate \cite{andrea6} and, in the context of rainbow devices, by using chirped graded array on a plate \cite{tian17a} and thin beams with tunable piezoelectric patches \cite{celli1}. %or in the context of gradient index (GRIN) lenses created by graded structuration to control elastic symmetric ($S_0$) and antisymmetric ($A_0$) waves  \cite{torrent_plate2} and to obtain deeply subwavelength focussing \cite{andrea4},  cloaking \cite{andrea5} and GRIN lenses (e.g. Luneburg, Maxwell-fisheye and Eaton \cite{luneburg_optic,celli1}) using resonator arrays \cite{andrea6}. 
A key simplification of thin elastic plates (or thin beams) is that the aspect ratio of plate thickness $b$ to wavelength $\lambda$ is limited to $b\ll \lambda$ and low frequency flexural ($A_0$) waves are dominant. %In contrast to deep elastic layers, both the elastic plate modelling, and that in acoustics, only involves a single scalar governing partial differential equation (PDE): 4th order Kirchhoff-Love PDE and 2nd order Helmholtz PDE respectively. 
This is not the case for deep elastic substrates that support both bulk compressional and shear waves \cite{achenbach84a}, with different wavespeeds, moreover both wavetypes couple and mode convert at interfaces with the Rayleigh surface wave \cite{rayleigh85a}. 
%This   his being the origin of many successful devices for surface imaging such as the acoustic microscope \cite{briggs09a} as well as the more well-known surface acoustic wave \cite{khelif_pillars1}
Because of the lack of this wave polarization in acoustics or flexural thin plates, recent experiments with acoustic waves \cite{acou_rain,sensing} in graded grooved waveguides and graded elastic plates \cite{tian17a,celli1} do not support mode conversion and can only demonstrate trapped rainbow phenomena. On the other hand, full elastic models, taking account of graded subwavelength resonator arrays have only just appeared  were recently envisaged at the low frequencies ($10 \sim 50 $ Hz) and  large-scale (tens or hundreds of metres) associated with geophysical applications \cite{colombi16a}. A key claim of this theoretical work is that one not only has elastic wave trapping and Rayleigh wave rainbow phenomena due to mechanisms described earlier, but one can in addition engineer mode conversion from surface to bulk modes. Furthermore, it is claimed that this mode conversion can create a surface array that is simultaneously reflectionless (as opposed to the rainbow trapping), yet has zero transmission into surface waves, and moreover that this occurs over an ultra-broadband range of frequencies. 

In the following, we move the entire theory from seismic to ultrasonic frequencies, and scales from decametres to millimetres, to design and build experiments that conclusively demonstrate both elastic Rayleigh wave rainbow trapping and mode conversion phenomena. We develop the physical interpretation of these observed results through fully 3D time dependent elastic numerical simulations and Bloch-Floquet theory for resonant elastic halfspaces \cite{daniel_andrea}. Finally we discuss some potential applications of Rayleigh trapping and conversion in the context of vibration absorption, signal filtering, energy harvesting and enhanced sensing. 

%The Rayleigh wave rainbow trapping has substantial implications in the design of energy concentrator devices (sensing, absorption and harvesting \cite{yan2013,Gonella2009621,tol_waveguide}) as we can strongly enhance the magnitude of the elastic signal selectively in space by frequency. Wave conversion is attractive in ground and mechanical vibration mitigation devices where converting Rayleigh waves, that typically travel long-distances with little attenuation and readily scatter from surface defects into bulk shear waves that can be attenuated by vibration black hole devices and other absorbers fabricated using the proposed graded design. 

\section*{Designing the graded subwavelength array}
\begin{figure}
\centering{}\includegraphics[clip,width=18cm,trim = 0mm 0mm 0mm 0mm]{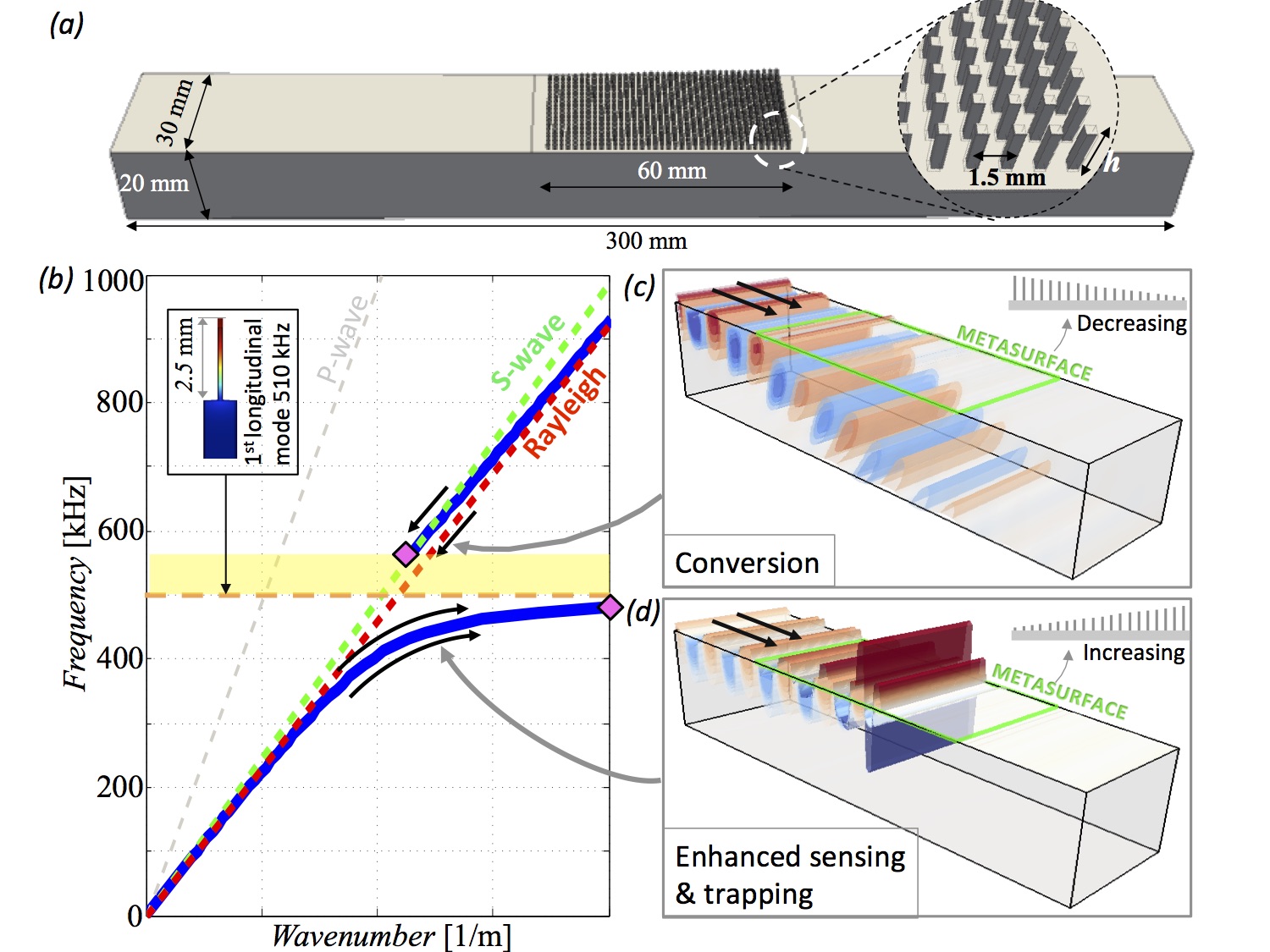}
\protect\caption{(a) Design and dimensions of the metasurface discussed in this study. (b) The dispersion curves of an elastic halfspace for ultrasonic frequencies measured at the top surface featuring surface Rayleigh (red) and shear (green) $S$-waves. The $P$ wave dispersion is marked with a dashed grey line. The introduction on the surface of local vertical resonators, (shown in the inset with the corresponding modal deformation), modifies the dispersion properties introducing a hybrid curve (blue) characterised by a bandgap (yellow area) bounded from below by an ultra slow mode and from above by a fast wave (pink diamonds). The $S$-wave serves as the light line, limiting the maximum speed of this system. (c) When Rayleigh waves approach an array of resonators of decreasing height (increasing resonance frequency), they are smoothly converted into $S$-waves. (d) Conversely, when the resonators are of increasing height, waves are spatially segregated, amplified and reflected. \label{fig:1}}
\end{figure}

Figure \ref{fig:1}a sketches the metasurface configurations: an aluminium block of rectangular cross-section (20-mm-thick, 40-mm-wide and 300-mm-long) with an array of decreasing or increasing resonators centred on the top surface; as  discussed in the methods section, the array of vertical rods is fabricated on top of the block using precision milling from a single metal block. The orientation of the graded array with the shortest, or longest, resonators facing the incident field determines whether Rayleigh wave trapping (Fig.~\ref{fig:1}c) or mode conversion occurs (Fig.~\ref{fig:1}d); the mechanical properties of aluminium are summarised in Tab.~\ref{tab:1}.  

To clearly interpret the physics, the characteristics of the metasurface are initially explored considering an infinite periodic array of resonators, of identical height  (see inset in Fig.~\ref{fig:1}b and rod's properties Tab.~\ref{tab:1}). Periodicity enables us to utilize Bloch theory to consider a single resonator in a cell with Floquet-Bloch condition upon the vertical edges of the cell. The resulting dispersion curve (blue line) relating phase shift across the cell to frequency is calculated using a 2D analytical approach that couples the longitudinal motion of the microrods with the full Navier elasticity equations in the halfspace \cite{daniel_andrea}. The resulting shape of the dispersion curve is characterised by ``hybridization" between the out-of-plane component ($u_z$) of the Rayleigh wave and the fundamental longitudinal mode in the rods \cite{fano,matthieu,PhysRevLett.111.036103}. Notably, in this framework, flexural modes within the rods have  negligible effect and can be ignored.
\begin{table}[h]
\centering
\begin{tabular}{ccccc}
\hline 
Height  & Section & Spacing & Young modulus & Density \tabularnewline
$h$ [mm]  & $A$ [mm$^2$]  & $s$ [mm] & $E$ [GPa]  & $\rho$ [kg/m$^3$] \tabularnewline
\hline 
2.5  & 0.025 & 1.5  & 69.0  & 2710.0 \tabularnewline
\hline 
\end{tabular}\protect\caption{Resonator parameters for Fig.~\ref{fig:1}b (constant height, cross-sectional area and spacing) and the mechanical properties of aluminium. \label{tab:1}}
\end{table} 

The dispersion properties of the metasurface strongly differ from that of the bare aluminium block for which we show the dispersionless sound-lines associated with compressional $P$-waves (grey dashed), shear $S$-waves (dashed green) and surface Rayleigh waves  (dashed red line). The resonance frequency (orange dashed) line punches through the sound-lines and, as is well-known, this perturbation leads to dispersion curve repulsion related to eigenvalue avoided crossing \cite{perkins86a,landau58a}, and the  creation of a band-gap (yellow shaded zone). It is clear that the array of constant height resonators is very powerful in terms of stopping (bandgaps, yellow shaded zone) and modifying the Rayleigh wave group velocity (flat branches for high wavenumbers). However the broadband performance is poor as control is achieved only around resonance that, in this case given the height of the aluminium microrod ($h$=2.5-mm), is located at $\sim 510$ kHz (dashed horizontal line). 
Graded metamaterial designs (e.g. Fig.~\ref{fig:1}c and d) overcome this limitation in bandgap size to create ultra-broadband devices by introducing more degrees of freedom to   engineering dispersion curves; since the resonators underpin the dispersion properties, their dynamic response can be tuned to build metamaterials with spatially graded properties \cite{andrea6}. For the microresonator considered here, the fundamental longitudinal mode frequency $f$ is defined as:
\begin{equation}
\label{eqn:risonanza}
f=\frac{1}{4 h}\sqrt{\frac{E}{\rho}},
\end{equation} 
 where $h$ is the resonator height, $E$ its Young's modulus and $\rho$ its density. 
In this formula longitudinal resonance is inversely proportional to the resonator height, $h$,  that is then chosen as the tuning parameter for the metasurface in Fig.~\ref{fig:1}c and d.
Despite the complexity added by the graded design, Eq.~(\ref{eqn:risonanza}), complemented with the dispersion diagram in Fig.~\ref{fig:1}b, is sufficient to fully describe the rainbow trapping  and to reveal a unique feature associated with elastic waves: the conversion of surface Rayleigh into bulk shear $S$-waves. Rainbow trapping is generated by the nearly flat branch approaching zero group velocity at the edge of the Brillouin zone, while the conversion is due to a hybrid mode, typical of this metasurface, connecting the Rayleigh with the $S$-wave line. Both phenomena are functions of the resonance frequency of the microrods and for the proposed designs in Fig.~\ref{fig:1}c and d, the trapping and turning points are precisely predicted using Eq.~(\ref{eqn:risonanza}) (see later discussions and Fig.~\ref{fig:4}). 

\begin{figure}
\centering{}\includegraphics[clip,width=18cm,trim = 0mm 0mm 0mm 0mm]{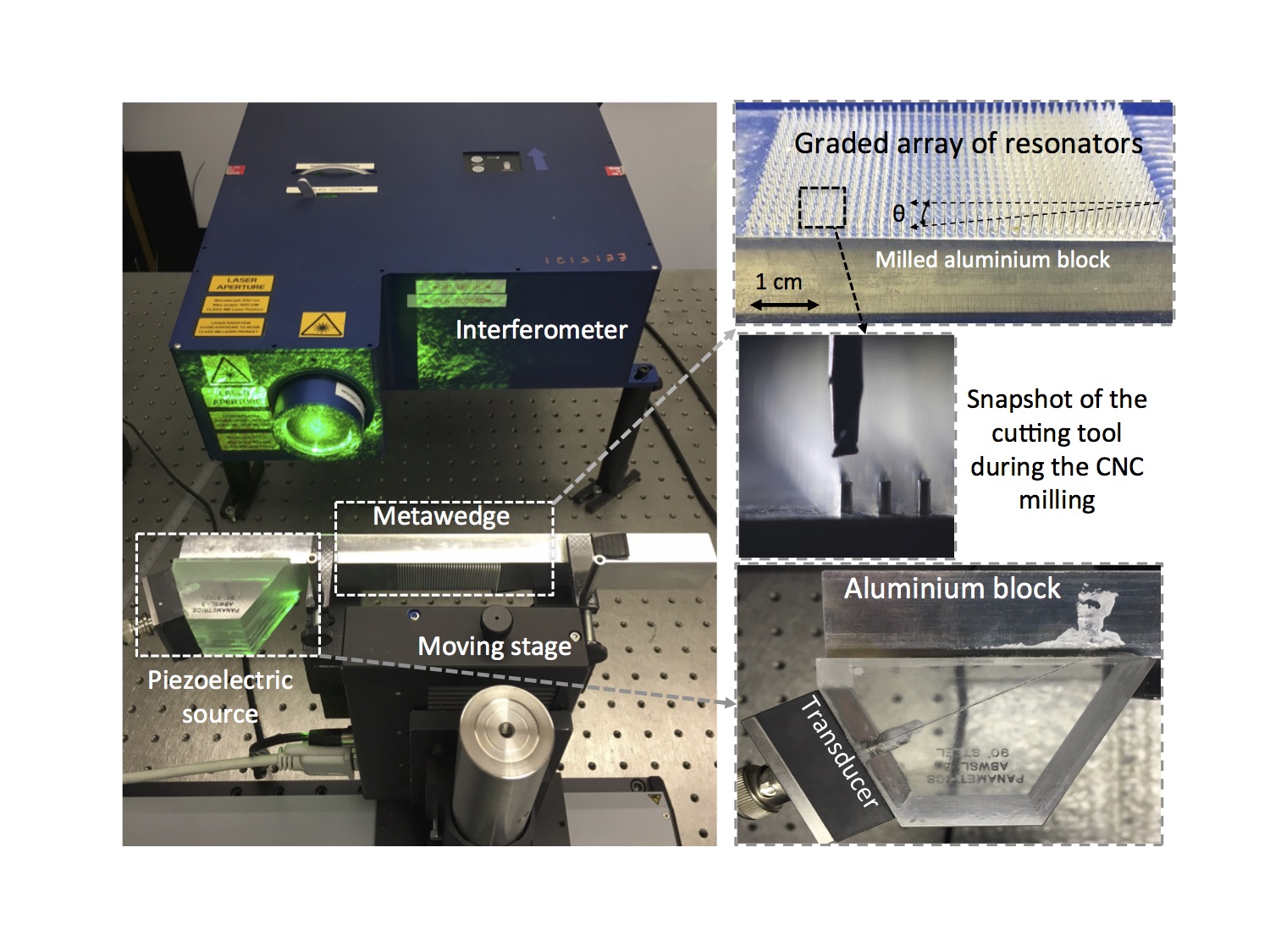}
\protect\caption{
The experimental set-up consisting of a laser interferometer scanning the surface of the metawedge and the aluminium block mounted on a moving stage. The excitation is obtained with an ultrasonic transducer coupled with a 65$^{\circ}$ wedge that, by phase matching, filters out all incident waves except Rayleigh waves. The resonators have a 0.5 $\times$ 0.5-mm cross-section and an increasing (or decreasing) height from 0.5-mm to 3-mm and have been obtained by micro-milling of the surface.  \label{fig:2}}
\end{figure}

Using the interpretation and modelling, the metawedge design chosen consists of a 1.5-mm spaced array of 40 rows each containing 18 microresonators with wedge height varying with an angle $\theta\sim 2.5^{\circ}$ (more details in Figs.~\ref{fig:1} and \ref{fig:2} and in the Methods section) from 3 mm the tallest to 0.5 mm the shortest.
The finite size of the aluminium block, and the presence of free surface boundaries at the surface of the block, leads to small reflections whose effect is quantified by analysing 3D numerical simulations of the experimental set-up implemented using a spectral element solver SPECFEM3D (see Methods).

\subsection*{Elastic rainbow and mode conversion}
Once fabricated, the metawedge was mounted on a moving stage as shown in Fig.~\ref{fig:2}; a laser adaptive photorefractive interferometer scans the surface of the block and records the displacement field in the out of plane direction $u_z$. An ultrasonic transducer generates, via phase matching with a $65^{\circ}$ wedge, purely Rayleigh waves at the surface of the aluminium block. The input signal consists of a 3 cycle sinusoid propagating towards the metasurface. The signal excites frequencies between 400 and 600 kHz with a maximum around 500 kHz (see further details in Fig.~\ref{fig:5}). The scanning procedure is repeated iteratively with the laser focal point moved around the block surface on a 2D grid (see Methods for details on the experimental set-up). To validate experimentally the Rayleigh to $S$ wave conversion the back surface is also scanned. Given the type of source and the experimental settings the surface displacement is typically of the order of a few nanometres.     

\begin{figure}
\centering{}\includegraphics[clip,width=18cm,trim = 0mm 0mm 0mm 0mm]{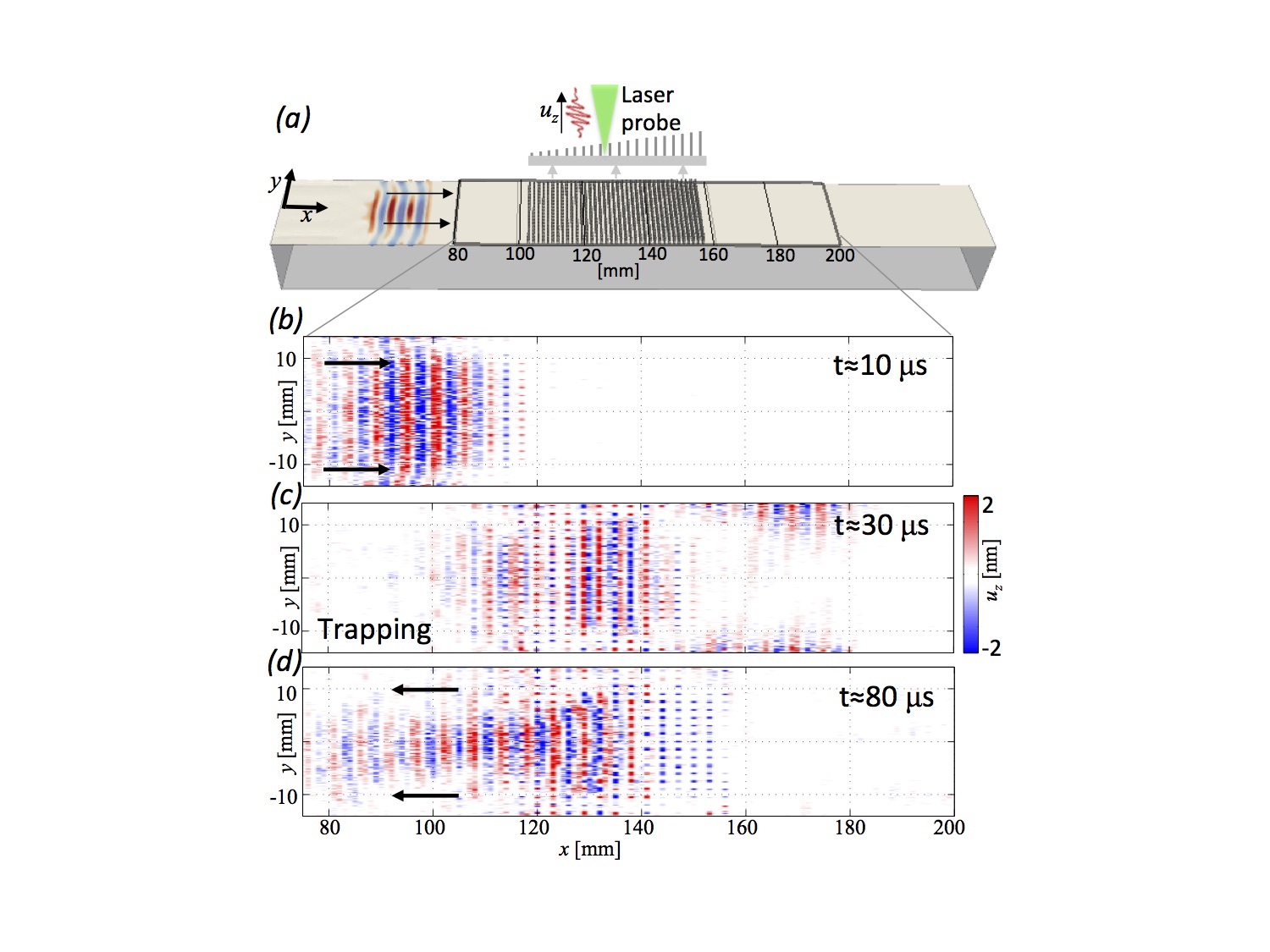}
\protect\caption{Experimental results for an increasing wedge profile. Data have been filtered between 450 and 650 kHz to improve the visualisation. (a) The numerical representation of the experiment. (b) The train of surface Rayleigh waves before entering the wedge, (c) trapped inside the wedge and (d) reflected backward. The underside of the sample was also scanned and no wave was found (see figure 4 and supplementary videos)
 \label{fig:3}}
\end{figure}

We demonstrated rainbow trapping by launching a Rayleigh wave at the metawedge from the short side (see the configuration in Fig.~3a). The train of Rayleigh waves approaches the metawedge interacting with the shortest resonators first and the interferometers scans the top surface. The snapshot in Fig.~3a, extracted from a 3D numerical simulation, shows that the plane wavefront is only marginally affected by the finite nature of the aluminium block and the Rayleigh waves primarily propagate as plane waves. 
The surface displacement scans in Fig.~\ref{fig:3}b-d show respectively the first interaction with the metasurface, the rainbow trapping and finally the reflected wavefield moving backward (video available as supplementary material). The underside of the sample was also scanned and no wave was found (see Fig.~\ref{fig:5}) The elastic rainbow effect spatially segregates the broadband signal based upon the frequency inside the metasurface, hence low frequency waves propagates longer in the metawedge while high frequency are trapper earlier. As a result of this frequency dependent phenomenon, the backward propagating field in Fig.~\ref{fig:3}d has lost the signal spatial coherence that characterised the input wavefield in Fig.~\ref{fig:3}b. Some energy leaks out of the metasurface because of waves guided by the lateral boundaries (Fig.~\ref{fig:3}c). Notably, during the trapping, the wave is strongly slowed down and the amplitude magnitude saturates the colorscale. 

The effectiveness of rainbow trapping, and in particular the spatial segregation by frequency, is analysed in Fig.~\ref{fig:5}a using experimental data. Scans of the wavefield are taken along the centreline of the top and the bottom surface and for each scanned position along the centreline a Fourier transform in time is calculated; the records are stopped before the main wavefront reaches the far right side to avoid spurious reflection. The analysis of the data focuses on the frequency band excited by the 3-cycles sinusoidal source. 
Fig.~\ref{fig:5}a, besides the metawedge configuration, shows the magnitude of the Fourier coefficients as a function of position and frequency for the top and bottom surface. As predicted for the increasing metawedge, the signal at the bottom surface remains close to the ambient noise level with the most interesting phenomena happening at the top surface. The rainbow trapping is visible in the surface plot where lower frequency signal travels much longer through the metawedge. The trapping point (or equally turning point in Fig.~\ref{fig:5}b), is highlighted by a marked increase in amplitude (dark shaded stripe) and is exactly predicted using Eq.~(\ref{eqn:risonanza}) combined with the metawedge geometry; the predicted trapping/ turning point is shown by the blue line. As a detail, another patch where amplification occurs, is visible before the trapping point and further analysis shows that it is created by one of the flexural resonances of the rod.
% this interpretation follows as the pattern differs from the longitudinal one because of the different dependence of the flexural modes with respect to $h$.
When the surface waves are slowed down inside the metawedge, their wavelength become as short as the spacing between the resonators. At the same time, the amplitude is strongly amplified at the surface and in the resonator. %During our experiment, we have recorded such a strong amplification that it has, in some cases, completely saturated the LDV. 
For further clarification of this amplification, the bottom plot of Fig.~\ref{fig:5}a, shows the magnitude of the temporal Fourier coefficient for a frequency of 500 kHz (the central frequency of the input source) at different positions along the surface. To exclude any possible local effects induced by the presence of the resonators on the scanned area, the blue line results from an average over 5 different measurements each taken a few millimetres away from the centreline. Before interacting with the wedge, the amplitude of the Fourier coefficients is typically between 70 and 80 pm. In the trapping phase, this value increases dramatically reaching, on top of the resonators, magnitude $\gg$ than 300 pm. On the right side of the wedge nothing is transmitted with the signal dropping below the noise threshold. Thus, besides the trapping, the graded design can realise ultrawide bandgaps.  

\begin{figure}
\centering{}\includegraphics[clip,width=18cm,trim = 0mm 0mm 0mm 0mm]{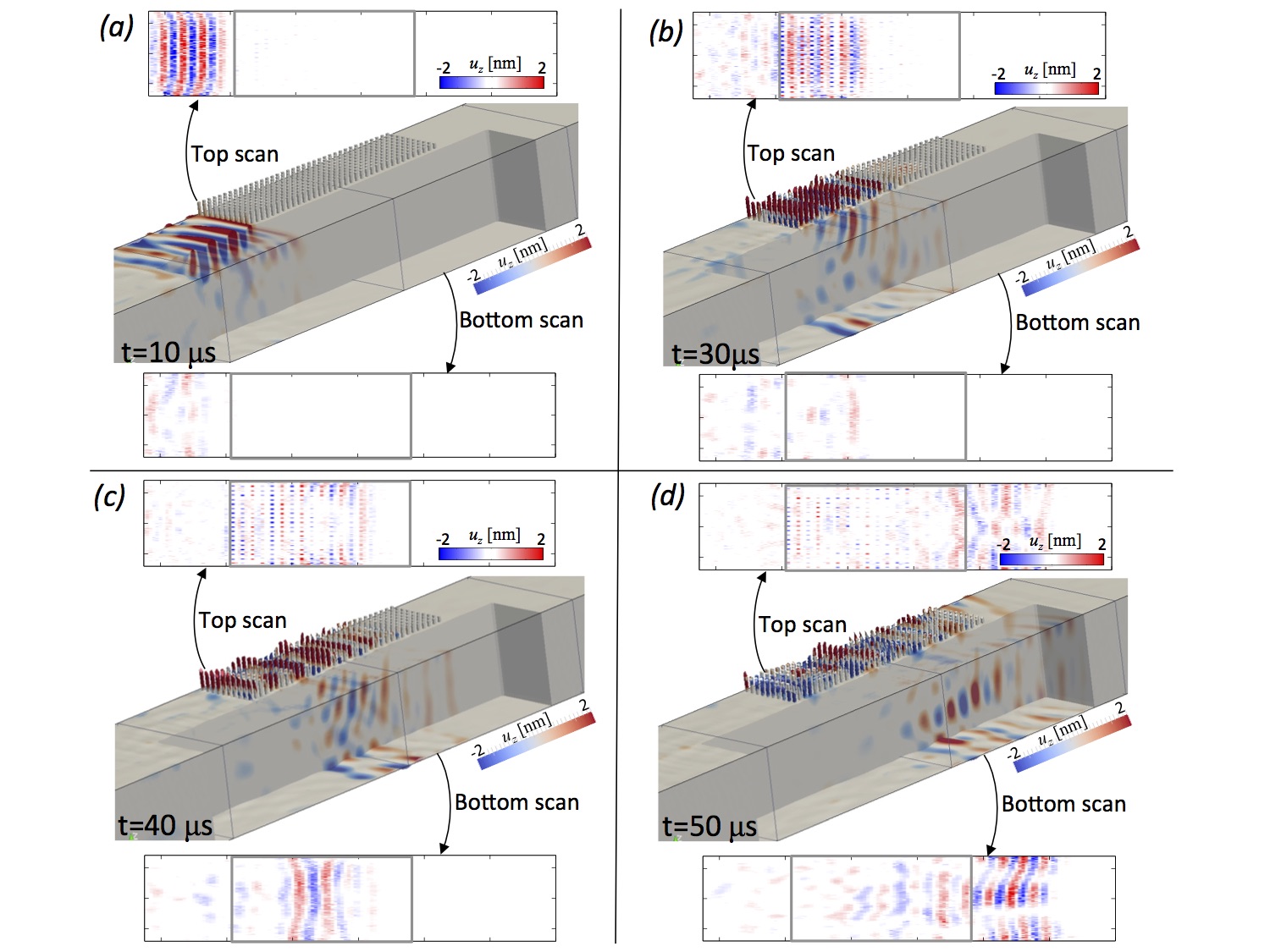}
\protect\caption{The conversion of surface Rayleigh into shear, $S$-waves seen both from the numerical and experimental perspective. (a) The top and bottom surface scans from the experiment are shown together  with the corresponding time-frame of a 3D numerical simulation reproducing the experiment. (b-d) same as (a) but for later time-steps.
 \label{fig:4}}
\end{figure}

\begin{figure}
\centering{}\includegraphics[clip,width=17cm,trim = 0mm 0mm 0mm 0mm]{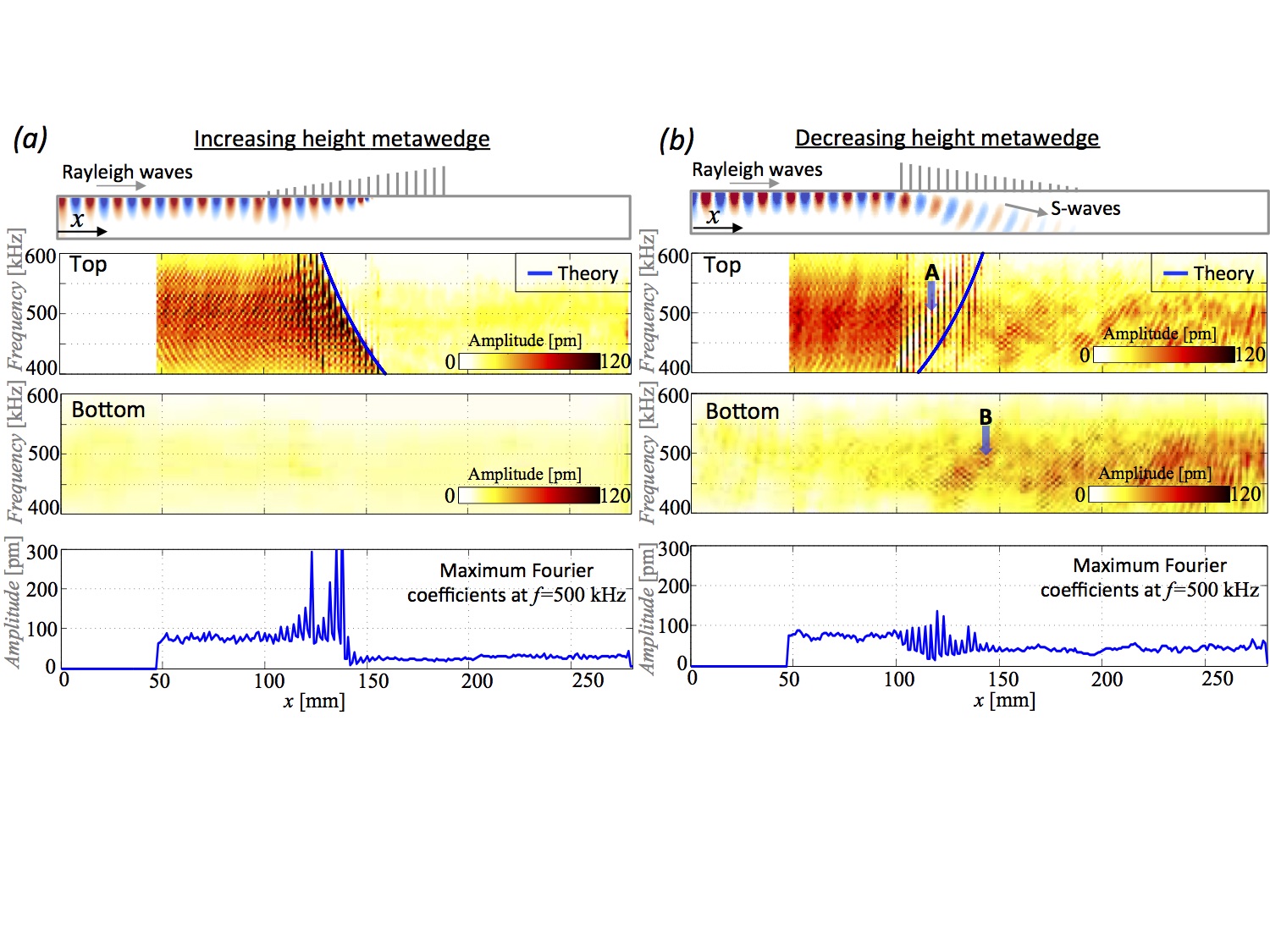}
\protect\caption{(a) Space-frequency analysis of the experimental data (displacement) for the increasing metawedge. From top to bottom: the geometry, and field. The next two panels show the power spectra along the $x$-direction in the aluminium block measured at the centreline for the top and bottom surfaces. The white area in the surface plot is due to the transducer. The lowest plot represents the maximum value of the Fourier coefficients at 500 kHz for the scan located approximately at the center of the top surface ($y$=0). (b) Same as (a) but for the inverse metawedge. The blue line in the top scan indicates the theoretical prediction of the turning point at the surface. 
 \label{fig:5}}
 
\end{figure}

Figure \ref{fig:4} reverses the orientation of the metawedge with the incident Rayleigh waves interacting with the longest resonators first, so the array is of decreasing height. To detect the conversion of Rayleigh into $S$-waves, the scanning procedure includes, as well as the top, also the bottom surface where the converted $S$-wave signal would be detected. To rule out any other boundary conversions that could create the signal  we complement the  experimental results with numerical simulations for the same propagation distance and geometry (videos are available as supplementary material). Figure~\ref{fig:4} shows experimental detail on both the top and bottom of the sample, as in the rainbow trapping one observes in Fig.~\ref{fig:4}a and b that there is an incoming Rayleigh wave and it interacts strongly with the resonators. Unlike for rainbow trapping the signal is not reversed but propagates downwards through the sample to impact upon the bottom surface where it is picked up by the bottom scan. To demonstrate that this is not some artificial effect or spurious side reflection we show the accompanying numerical simulation in each panel with Fig.~\ref{fig:4}c showing the surface wave having been deflected in the $S$-wave. Fig.~\ref{fig:4}d shows the bottom scan having picked up a surface wave along this lower surface of the block and some signal having been reflected and impacting back upon the top scan; these can be attributed to the finite nature of the sample which is verified numerically. 

We now move to Fig.~\ref{fig:5}b where the generation of $S$-waves is analysed in detail with the same method used for the increasing metawadge. The frequency-position plot for the bottom scan, in this case, shows strong signals produced by the converted Rayleigh waves; the turning point is identified from the top scan and fits with the theoretical prediction. Because of the graded design, the transition between Rayleigh to $S$-waves happens very smoothly, without back reflection \cite{colombi16a}, as opposed to Fig.~\ref{fig:5}a where the strong bandgap effect rules the behaviour and does not allow the wave to propagate further. The turning point is still identified from Eq. (\ref{eqn:risonanza}), but the prediction is slightly less accurate because the conversion happens at the upper limit of the bandgap, and not exactly at the resonance, in contrast to the trapping. The prediction can be made accurate through a complex formula for the conversion given in a previous work\cite{daniel_andrea,colombi16a}. Nevertheless the physics emerges clearly such that the refraction angle generated by the conversion and measured between point A and B in Fig.~\ref{fig:5}b matches the Snell's law prediction for Rayleigh to shear wave conversion in aluminium.
As suggested by the Fourier coefficient at 500 kHz plot at the bottom in Fig.~\ref{fig:5}b, the conversion happens smoothly, with no clear sign of amplification (only small oscillations due to the rod's resonance are visible), nor back reflection.
After the conversion, the signal bounces up and down between the top and bottom surfaces as it is reasonable to expect given the geometry of the aluminium block.

\section*{Discussion}

We have conclusively demonstrated via experiments in the ultrasonic regime, complemented by theory and numerical simulation, that mechanical metasurfaces can be designed to create elastic rainbow trapping and to generate surface Rayleigh wave to bulk shear wave conversion. Both effects have great potential to achieve performance and functionalities hitherto unachievable. 
Because local resonance is at the origin of the elastic rainbow and the mode conversion, this technology can be tuned to the desired lengthscale through an appropriate engineering. Potential applications include, but are not limited to, resonant seismic barriers, structural vibration mitigation, micro and nano electromechanical components and the field of waveguiding of ultra and hypersounds.    
For instance the rainbow trapping leads to strong spatial localisation, segregated by frequency, of vibration energy ideally suited to energy harvesting applications and we envisage this opening up new possibilities. The waste of energy through vibration, and the potential to harvest it and generate efficiency savings is substantial. 
In addition, the spatial segregation by frequency is suggestive that this can be used as a wave filter, a shorter wedge removing all frequencies within a band, and as a wave sensor to pick out specific components of the wave spectrum. 
These ideas are closely related to the acoustic black hole concept where an area with exponentially increasing refractive index ($>1$) attracts and ultimately traps waves \cite{krylov_black_holes,Anderson2015141}. In practice, this technology can be applied only to elastic plates with applications similar to those discussed here: With the introduction of the metawedge, a resonant elastic black hole can be envisaged. 
%Energy focusing, frequency selectivity and harvesting, when combined together, may be used to build new self-powered elastic wave sensors, an active research topic of continuous structural health monitoring.  
%Concentrators and energy harvesters from structured media are still based on phononic crystals and plates with chirped structures \cite{tol_waveguide,yan2013} or materials with very complex geometries that are difficult to fabricate \cite{Gonella2009621}.  
%
Mode conversion, allowing surface waves to be diverted into bulk waves, is also a powerful concept creating the existence of regions of surface wave silence or equivalently protection from surface waves; this has clear application to many vibration related problems in mechanical engineering (high speed and high precision manufacturing), civil engineering and geophysics where the area of seismic metamaterials and seismic protection devices is very topical \cite{andrea_tree,Cacciola201547}.

\section*{Methods}
\subsection*{Metawedge fabrication}
The sample was CNC milled out of a solid block of aluminium using a 1mm diameter cutter (machining time: 22 h). The finished sample was approximately 300-mm $\times$ 30-mm $\times$ 20-mm.
The metawedge structure consists of resonators with 0.5$\times$0.5-mm cross-section and of heights graded from 0.5-mm to 3-mm over 60-mm along the block and spaced by $~$1.5-mm placed midway along the block and covered the full width of the block.
The width of the block was chosen so that the ultrasound could be excited reasonably uniformly across the full width.  The thickness of the block (20-mm) was chosen to be thick enough to render any dispersion (due to Lamb waves in the finite block) irrelevant so the surface wave could reasonably be treated as a Rayleigh wave at the frequency used in the experiment (500 kHz / $\lambda\sim6$-mm).
The length of the block was chosen so there was ample space to fit the transducer used at either end of the block and differentiate all the separate echoes and signals received in time.  The sample was milled flat on the top and bottom surfaces and held in a jig that allowed the ultrasound to be observed either on the top or bottom surface and with the ultrasound propagating from either end of the block.

\subsection*{Experimental set-up and measurements}
The sample surface was smoothly milled but optically rough and a rough-surface capable optical detector (Bossanova Tempo-10HF) was used as a detector. The detector was used in absolute calibration mode so that the signal amplitude measured at each point was directly proportional to the surface amplitude regardless of the optical return at that point.
The ultrasound was excited using a Panametrics Videoscan V414 0.5 MHz plane wave transducer with a $65^{\circ}$ polymer wedge to couple the longitudinal wave of the transducer into a Rayleigh wave on the sample. The transducer and wedge were glued to the sample using phenyl salicylate which provided good coupling and long term stability while allowing easy removal and reattachment of the transducer.
The transducer was driven by a Ritec RPR-4000 programmable pulser using a 3 cycles sinusoidal burst at 500 kHz with an amplitude of 195 V peak-to-peak and repetition rate of 500 Hz.  At this repetition rate, all echoes from the previous pulse were able to die out before the next measurement.
The signal was captured using a digital storage oscilloscope (Lecroy Waverunner104Xi-A) and averaged 100 times before transfer to a desktop computer. 
The sample was mounted on scanning stages with a movement range of 300-mm $\times$ 50-mm and nominal positional accuracy of 1-$\mu$m allowing the full surface of the sample to be scanned. Scanning was performed at a rate of $\sim$2 points/second.  This scan rate was primarily limited by the settling time of the optical detector between positions on the sample surface.

\subsection*{Numerical simulations}
The propagation of surface waves in a 3D halfspace is a common 
problem in numerical elastodynamics and modeled applying 
traction free conditions on the top-surface and on the micropillars. In order to compare the simulations quantitatively to the experimental results, traction free conditions are also applied to the sides of the aluminium block in the simulation thereby representing the aluminum block sides of the experimental sample. Only the side closest to the source is supplied with absorbing boundaries (perfectly matched layers \cite{Komatitsch_CPML}) to suppress the backward reflection of the Rayleigh wave. In the actual experiment, this reflection is also not present because the plastic wedge between the transducer and the aluminium block (Fig.~\ref{fig:2}) only allows Rayleigh waves to propagate forward towards the metasurface.
As in the experiment, the plane Rayleigh wave front is obtained with a 3 cycle sinusoidal  source time function and 40 point source synchronously triggered on the top surface at a distance of 6 cm from the first row of resonators.
The 3D time domain simulations have been carried out using SPECFEM3D a code that solves the elastic wave equation using finite difference in time (Newmark scheme) and the spectral element method in space ($5^{th}$ order polynomials) \cite{specfem_cart}. The parallelization is implemented through domain decomposition with MPI and the mesh is made of hexahedra elements and it is generated using the commercial meshing software CUBIT. Given the high quality factor of aluminium ($\gg$ 1000), attenuation has not been considered.
Simulations are then run on a parallel cluster (Froggy at University of Grenoble) on 256 CPUs. 3D plots have been generated with Paraview while 2D with Matlab and Matplotlib. 
%

%\bibliographystyle{Science}
%\bibliography{scibib}%,biblio_meta}

\begin{thebibliography}{10}
\expandafter\ifx\csname url\endcsname\relax
  \def\url#1{\texttt{#1}}\fi
\expandafter\ifx\csname urlprefix\endcsname\relax\def\urlprefix{URL }\fi
\providecommand{\bibinfo}[2]{#2}
\providecommand{\eprint}[2][]{\url{#2}}

\bibitem{colombi16a}
\bibinfo{author}{Colombi, A.}, \bibinfo{author}{Colquitt, D.},
  \bibinfo{author}{Roux, P.}, \bibinfo{author}{Guenneau, S.} \&
  \bibinfo{author}{Craster, R.~V.}
\newblock \bibinfo{title}{A seismic metamaterial: The resonant metawedge}.
\newblock \emph{\bibinfo{journal}{Sci. Rep.}} \textbf{\bibinfo{volume}{6}},
  \bibinfo{pages}{27717} (\bibinfo{year}{2016}).

\bibitem{acou_rain}
\bibinfo{author}{Zhu, J.} \emph{et~al.}
\newblock \bibinfo{title}{Acoustic rainbow trapping}.
\newblock \emph{\bibinfo{journal}{Sci. Rep.}} \textbf{\bibinfo{volume}{3}},
  \bibinfo{pages}{1728} (\bibinfo{year}{2013}).

\bibitem{romero13a}
\bibinfo{author}{Romero-Garcia, V.}, \bibinfo{author}{Pico, R.},
  \bibinfo{author}{Cebrecos, A.}, \bibinfo{author}{Sanchez-Morcillo, V.~J.} \&
  \bibinfo{author}{Staliunas, K.}
\newblock \bibinfo{title}{Enhancement of sound in chirped sonic crystals}.
\newblock \emph{\bibinfo{journal}{Appl. Phys. Lett.}}
  \textbf{\bibinfo{volume}{102}}, \bibinfo{pages}{091906}
  (\bibinfo{year}{2013}).

\bibitem{hess2007}
\bibinfo{author}{Tsakmakidis, K.~L.}, \bibinfo{author}{Boardman, A.~D.} \&
  \bibinfo{author}{Hess, O.}
\newblock \bibinfo{title}{{Trapped rainbow} storage of light in metamaterials}.
\newblock \emph{\bibinfo{journal}{Nature}} \textbf{\bibinfo{volume}{450}},
  \bibinfo{pages}{397--401} (\bibinfo{year}{2007}).

\bibitem{li}
\bibinfo{author}{Li, D.}, \bibinfo{author}{Zigoneanu, L.},
  \bibinfo{author}{Popa, B.-I.} \& \bibinfo{author}{Cummer, S.~A.}
\newblock \bibinfo{title}{Design of an acoustic metamaterial lens using genetic
  algorithms}.
\newblock \emph{\bibinfo{journal}{The Journal of the Acoustical Society of
  America}} \textbf{\bibinfo{volume}{132}}, \bibinfo{pages}{2823--2833}
  (\bibinfo{year}{2012}).

\bibitem{matthieu}
\bibinfo{author}{Rupin, M.}, \bibinfo{author}{Lemoult, F.},
  \bibinfo{author}{Lerosey, G.} \& \bibinfo{author}{Roux, P.}
\newblock \bibinfo{title}{Experimental demonstration of ordered and disordered
  multi-resonant metamaterials for {L}amb waves}.
\newblock \emph{\bibinfo{journal}{Phys. Rev. Lett.}}
  \textbf{\bibinfo{volume}{112}}, \bibinfo{pages}{234301}
  (\bibinfo{year}{2014}).

\bibitem{andrea4}
\bibinfo{author}{Colombi, A.}, \bibinfo{author}{Roux, P.} \&
  \bibinfo{author}{Rupin, M.}
\newblock \bibinfo{title}{Sub-wavelength energy trapping of elastic waves in a
  meta-material}.
\newblock \emph{\bibinfo{journal}{J.\ Acoust.\ Soc.\ Am.}}
  \textbf{\bibinfo{volume}{136}}, \bibinfo{pages}{EL192--8}
  (\bibinfo{year}{2014}).

\bibitem{kaina}
\bibinfo{author}{Kaina, N.}, \bibinfo{author}{Lemoult, F.},
  \bibinfo{author}{Fink, M.} \& \bibinfo{author}{Lerosey, G.}
\newblock \bibinfo{title}{Ultra small mode volume defect cavities in spatially
  ordered and disordered metamaterials}.
\newblock \emph{\bibinfo{journal}{Appl. Phys. Lett.}}
  \textbf{\bibinfo{volume}{102}}, \bibinfo{pages}{--} (\bibinfo{year}{2013}).

\bibitem{yu14a}
\bibinfo{author}{Yu, N.} \& \bibinfo{author}{Capasso, F.}
\newblock \bibinfo{title}{Flat optics with designer metasurfaces}.
\newblock \emph{\bibinfo{journal}{Nature Materials}}
  \textbf{\bibinfo{volume}{13}}, \bibinfo{pages}{139--149}
  (\bibinfo{year}{2014}).

\bibitem{maigyte13a}
\bibinfo{author}{Maigyte, L.} \emph{et~al.}
\newblock \bibinfo{title}{Flat lensing in the visible frequency range by
  woodpile photonic crystals}.
\newblock \emph{\bibinfo{journal}{Optics Letters}}
  \textbf{\bibinfo{volume}{38}}, \bibinfo{pages}{2376--2378}
  (\bibinfo{year}{2013}).

\bibitem{kadic11a}
\bibinfo{author}{Kadic, M.}, \bibinfo{author}{Guenneau, S.},
  \bibinfo{author}{Enoch, S.} \& \bibinfo{author}{Ramakrishna, S.}
\newblock \bibinfo{title}{Plasmonic space folding: {F}ocussing surface plasmons
  via negative refraction in complementary media}.
\newblock \emph{\bibinfo{journal}{ACS Nano}} \textbf{\bibinfo{volume}{5}},
  \bibinfo{pages}{6819--6825} (\bibinfo{year}{2011}).

\bibitem{christensen12a}
\bibinfo{author}{Christensen, J.} \& \bibinfo{author}{{de Abajo}, F. J.~G.}
\newblock \bibinfo{title}{Anisotropic metamaterials for full control of
  acoustic waves}.
\newblock \emph{\bibinfo{journal}{Phys. Rev. Lett.}}
  \textbf{\bibinfo{volume}{108}}, \bibinfo{pages}{124301}
  (\bibinfo{year}{2012}).

\bibitem{lens_nature}
\bibinfo{author}{Kundtz, N.} \& \bibinfo{author}{Smith, D.~R.}
\newblock \bibinfo{title}{Extreme-angle broadband metamaterial lens}.
\newblock \emph{\bibinfo{journal}{Nat. Mater}} \textbf{\bibinfo{volume}{9}},
  \bibinfo{pages}{129--132} (\bibinfo{year}{2010}).

\bibitem{andrea6}
\bibinfo{author}{Colombi, A.}
\newblock \bibinfo{title}{Resonant metalenses for flexural waves}.
\newblock \emph{\bibinfo{journal}{J.\ Acoust.\ Soc.\ Am.}}
  \textbf{\bibinfo{volume}{140}}, \bibinfo{pages}{EL423}
  (\bibinfo{year}{2016}).

\bibitem{yan2013}
\bibinfo{author}{Yan, X.}, \bibinfo{author}{Zhu, R.}, \bibinfo{author}{Huang,
  G.} \& \bibinfo{author}{Yuan, F.-G.}
\newblock \bibinfo{title}{Focusing guided waves using surface bonded elastic
  metamaterials}.
\newblock \emph{\bibinfo{journal}{Applied Physics Letters}}
  \textbf{\bibinfo{volume}{103}}, \bibinfo{pages}{121901}
  (\bibinfo{year}{2013}).

\bibitem{torrent_plate2}
\bibinfo{author}{Jin, Y.}, \bibinfo{author}{Torrent, D.},
  \bibinfo{author}{Pennec, Y.}, \bibinfo{author}{Pan, Y.} \&
  \bibinfo{author}{Djafari-Rouhani, B.}
\newblock \bibinfo{title}{Simultaneous control of the {S0} and {A0} {L}amb
  modes by graded phononic crystal plates}.
\newblock \emph{\bibinfo{journal}{J. Appl. Phys.}}
  \textbf{\bibinfo{volume}{117}}, \bibinfo{pages}{244904}
  (\bibinfo{year}{2015}).

\bibitem{andrea_soil_lenses}
\bibinfo{author}{Colombi, A.}, \bibinfo{author}{Guenneau, S.},
  \bibinfo{author}{Roux, P.} \& \bibinfo{author}{Craster, R.}
\newblock \bibinfo{title}{Transformation seismology: composite soil lenses for
  steering surface elastic rayleigh waves}.
\newblock \emph{\bibinfo{journal}{Sci. Rep.}} \textbf{\bibinfo{volume}{6}},
  \bibinfo{pages}{25320} (\bibinfo{year}{2016}).

\bibitem{tian17a}
\bibinfo{author}{Tian, Z.} \& \bibinfo{author}{Yu, L.}
\newblock \bibinfo{title}{Rainbow trapping of ultrasonic guided waves in
  chirped phononic crystal plates}.
\newblock \emph{\bibinfo{journal}{Sci. Rep.}} \textbf{\bibinfo{volume}{7}},
  \bibinfo{pages}{40004} (\bibinfo{year}{2017}).

\bibitem{jang11a}
\bibinfo{author}{Jang, M.} \& \bibinfo{author}{Atwater, H.}
\newblock \bibinfo{title}{Plasmonic rainbow trapping structures for light
  localization and spectrum splitting}.
\newblock \emph{\bibinfo{journal}{Phys. Rev. Lett.}}
  \textbf{\bibinfo{volume}{107}}, \bibinfo{pages}{207401}
  (\bibinfo{year}{2011}).

\bibitem{gan2009}
\bibinfo{author}{Gan, Q.}, \bibinfo{author}{Ding, Y.} \&
  \bibinfo{author}{Bartoli, F.~J.}
\newblock \bibinfo{title}{{`Rainbow'} trapping and releasing at telecom
  wavelengths}.
\newblock \emph{\bibinfo{journal}{Phys. Rev. Lett.}}
  \textbf{\bibinfo{volume}{102}}, \bibinfo{pages}{056801}
  (\bibinfo{year}{2009}).

\bibitem{jimenez16a}
\bibinfo{author}{Jimenez, N.} \emph{et~al.}
\newblock \bibinfo{title}{Broadband quasi perfect absorption using chirped
  multi-layer porous materials}.
\newblock \emph{\bibinfo{journal}{AIP Advances}} \textbf{\bibinfo{volume}{6}},
  \bibinfo{pages}{121605} (\bibinfo{year}{2016}).

\bibitem{sensing}
\bibinfo{author}{Chen, Y.}, \bibinfo{author}{Liu, H.}, \bibinfo{author}{Reilly,
  M.}, \bibinfo{author}{Bae, H.} \& \bibinfo{author}{Yu, M.}
\newblock \bibinfo{title}{Enhanced acoustic sensing through wave compression
  and pressure amplification in anisotropic metamaterials}.
\newblock \emph{\bibinfo{journal}{Nat Comm.}} \textbf{\bibinfo{volume}{5}},
  \bibinfo{pages}{5247} (\bibinfo{year}{2014}).

\bibitem{tyc2009}
\bibinfo{author}{Leonhardt, U.} \& \bibinfo{author}{Tyc, T.}
\newblock \bibinfo{title}{Broadband invisibility by non-euclidean cloaking}.
\newblock \emph{\bibinfo{journal}{Science}} \textbf{\bibinfo{volume}{323}},
  \bibinfo{pages}{110--112} (\bibinfo{year}{2009}).

\bibitem{celli1}
\bibinfo{author}{Cardella, D.}, \bibinfo{author}{Celli, P.} \&
  \bibinfo{author}{Gonella, S.}
\newblock \bibinfo{title}{Manipulating waves by distilling frequencies: a
  tunable shunt-enabled rainbow trap}.
\newblock \emph{\bibinfo{journal}{Smart Materials and Structures}}
  \textbf{\bibinfo{volume}{25}}, \bibinfo{pages}{085017}
  (\bibinfo{year}{2016}).

\bibitem{achenbach84a}
\bibinfo{author}{Achenbach, J.~D.}
\newblock \emph{\bibinfo{title}{Wave propagation in elastic solids}}
  (\bibinfo{publisher}{Amsterdam: North-Holland.}, \bibinfo{year}{1984}).

\bibitem{rayleigh85a}
\bibinfo{author}{{Lord Rayleigh}}.
\newblock \bibinfo{title}{On waves propagated along the plane surface of an
  elastic solid}.
\newblock \emph{\bibinfo{journal}{Proc Lond Math Soc}}
  \textbf{\bibinfo{volume}{17}}, \bibinfo{pages}{4--11} (\bibinfo{year}{1885}).

\bibitem{daniel_andrea}
\bibinfo{author}{Colquitt, D.}, \bibinfo{author}{Colombi, A.},
  \bibinfo{author}{Craster, R.}, \bibinfo{author}{Roux, P.} \&
  \bibinfo{author}{Guenneau, S.}
\newblock \bibinfo{title}{Seismic metasurfaces: Sub-wavelength resonators and
  rayleigh wave interaction}.
\newblock \emph{\bibinfo{journal}{J. Mech. Phys. Solids}}
  \textbf{\bibinfo{volume}{99}}, \bibinfo{pages}{379--393}
  (\bibinfo{year}{2017}).

\bibitem{fano}
\bibinfo{author}{Miroshnichenko, A.~E.}, \bibinfo{author}{Flach, S.} \&
  \bibinfo{author}{Kivshar, Y.~S.}
\newblock \bibinfo{title}{Fano resonances in nanoscale structures}.
\newblock \emph{\bibinfo{journal}{Rev. Mod. Phys.}}
  \textbf{\bibinfo{volume}{82}}, \bibinfo{pages}{2257--2298}
  (\bibinfo{year}{2010}).

\bibitem{PhysRevLett.111.036103}
\bibinfo{author}{Boechler, N.} \emph{et~al.}
\newblock \bibinfo{title}{Interaction of a contact resonance of microspheres
  with surface acoustic waves}.
\newblock \emph{\bibinfo{journal}{Phys. Rev. Lett.}}
  \textbf{\bibinfo{volume}{111}}, \bibinfo{pages}{036103}
  (\bibinfo{year}{2013}).

\bibitem{perkins86a}
\bibinfo{author}{Perkins, N.~C.} \& \bibinfo{author}{Mote, C.~D.}
\newblock \bibinfo{title}{Comments on curve veering in eigenvalue problems}.
\newblock \emph{\bibinfo{journal}{J. Sound Vib.}}
  \textbf{\bibinfo{volume}{106}}, \bibinfo{pages}{451--463}
  (\bibinfo{year}{1986}).

\bibitem{landau58a}
\bibinfo{author}{Landau, L.~D.} \& \bibinfo{author}{Lifshitz, E.~M.}
\newblock \emph{\bibinfo{title}{Quantum Mechanics non-relativistic theory}}
  (\bibinfo{publisher}{Pergamon Press}, \bibinfo{year}{1958}).

\bibitem{krylov_black_holes}
\bibinfo{author}{Krylov, V.}
\newblock \bibinfo{title}{Acoustic black holes: recent developments in the
  theory and applications}.
\newblock \emph{\bibinfo{journal}{IEEE Transactions on Ultrasonics,
  Ferroelectrics, and Frequency Control}} \textbf{\bibinfo{volume}{61}},
  \bibinfo{pages}{1296--1306} (\bibinfo{year}{2014}).

\bibitem{Anderson2015141}
\bibinfo{author}{Anderson, B.~E.}, \bibinfo{author}{Remillieux, M.~C.},
  \bibinfo{author}{Bas, P.-Y.~L.}, \bibinfo{author}{Ulrich, T.} \&
  \bibinfo{author}{Pieczonka, L.}
\newblock \bibinfo{title}{Ultrasonic radiation from wedges of cubic profile:
  Experimental results}.
\newblock \emph{\bibinfo{journal}{Ultrasonics}} \textbf{\bibinfo{volume}{63}},
  \bibinfo{pages}{141 -- 146} (\bibinfo{year}{2015}).

\bibitem{andrea_tree}
\bibinfo{author}{Colombi, A.}, \bibinfo{author}{Roux, P.},
  \bibinfo{author}{Guenneau, S.}, \bibinfo{author}{Gueguen, P.} \&
  \bibinfo{author}{Craster, R.}
\newblock \bibinfo{title}{Forests as a natural seismic metamaterial: Rayleigh
  wave bandgaps induced by local resonances}.
\newblock \emph{\bibinfo{journal}{Sci. Rep.}} \textbf{\bibinfo{volume}{5}},
  \bibinfo{pages}{19238} (\bibinfo{year}{2016}).

\bibitem{Cacciola201547}
\bibinfo{author}{Cacciola, P.}, \bibinfo{author}{Espinosa, M.~G.} \&
  \bibinfo{author}{Tombari, A.}
\newblock \bibinfo{title}{Vibration control of piled-structures through
  structure-soil-structure-interaction}.
\newblock \emph{\bibinfo{journal}{Soil Dynamics and Earthquake Engineering}}
  \textbf{\bibinfo{volume}{77}}, \bibinfo{pages}{47 -- 57}
  (\bibinfo{year}{2015}).

\bibitem{Komatitsch_CPML}
\bibinfo{author}{Komatitsch, D.} \& \bibinfo{author}{Martin, R.}
\newblock \bibinfo{title}{An unsplit convolutional perfectly matched layer
  improved at grazing incidence for the seismic wave equation}.
\newblock \emph{\bibinfo{journal}{Geophysics}} \textbf{\bibinfo{volume}{72}},
  \bibinfo{pages}{SM155--SM167} (\bibinfo{year}{2007}).

\bibitem{specfem_cart}
\bibinfo{author}{Peter, D.} \emph{et~al.}
\newblock \bibinfo{title}{Forward and adjoint simulations of seismic wave
  propagation on fully unstructured hexahedral meshes}.
\newblock \emph{\bibinfo{journal}{Geophys. J. Int.}}
  \textbf{\bibinfo{volume}{186}}, \bibinfo{pages}{721--739}
  (\bibinfo{year}{2011}).

\end{thebibliography}

\section*{Acknowledgements}
All of the computations presented in this paper were performed using the Froggy platform of the CIMENT infrastructure (\textit{https://ciment.ujf-grenoble.fr}), supported by the Rhone-Alpes region (GRANT CPER07$\_$13 CIRA), the OSUG2020 labex (reference ANR10 LABX56) and the EquipMeso project (reference ANR-10-EQPX-29-01) of the programme Investissements d'Avenir supervised by the Agence Nationale pour la Recherche.
A.C and R.C. thanks the EPSRC for their support through research grants EP/I018948/1, EP/L024926/1, EP/J009636/1 and Mathematics Platform grant EP/I019111/1. A.C. was supported by the Marie Curie Fellowship ''Metacloak". A.C., P.R., S.G and R.C. acknowledge the  support of the French project Metaforet (reference ANR) that facilitates the collaboration between Imperial College, ISTerre and Institut Fresnel.

\section*{Author contributions}
A.C. and M.C. initiated the project, A.C. carried out the simulations; A.C. and D.C. derived the analytical form of the dispersion equation, V.A., R.J.S., R.P. and M.C. designed the ultrasonic experiment, Ad. C. fabricated the metawedge. P.R and R.C. helped with data analysis and the interpretation of results; A.C. created the figures. All authors have equally contributed to writing and editing.  

\section*{Additional information}
\subsection*{Competing financial interests} The authors declare no competing financial interests.


\end{document}